\documentclass{emulateapj}

\shorttitle{CANGAROO-III Observations of MSH\,15$-$52}
\shortauthors{Nakamori et al.}

\begin{document}
\title{Observation of an extended VHE gamma-ray emission from MSH\,15$-$52 with CANGAROO-III}
\author{
T.~Nakamori\altaffilmark{1}
H.~Kubo\altaffilmark{1}
T.~Yoshida\altaffilmark{2}
T.~Tanimori\altaffilmark{1}
R.~Enomoto\altaffilmark{3}
G.~V.~Bicknell\altaffilmark{4}
R.~W.~Clay\altaffilmark{5}
P.~G.~Edwards\altaffilmark{6}
S.~Gunji\altaffilmark{7}
S.~Hara\altaffilmark{8}
T.~Hara\altaffilmark{9}
T.~Hattori\altaffilmark{10}
S.~Hayashi\altaffilmark{11}
Y.~Higashi\altaffilmark{1}
Y.~Hirai\altaffilmark{2}
K.~Inoue\altaffilmark{7}
S.~Kabuki\altaffilmark{1}
F.~Kajino\altaffilmark{11}
H.~Katagiri\altaffilmark{12}
A.~Kawachi\altaffilmark{10}
T.~Kifune\altaffilmark{3}
R.~Kiuchi\altaffilmark{3}
J.~Kushida\altaffilmark{10}
Y.~Matsubara\altaffilmark{13}
T.~Mizukami\altaffilmark{1}
Y.~Mizumoto\altaffilmark{14}
R.~Mizuniwa\altaffilmark{10}
M.~Mori\altaffilmark{3}
H.~Muraishi\altaffilmark{15}
Y.~Muraki\altaffilmark{13}
T.~Naito\altaffilmark{9}
S.~Nakano\altaffilmark{1}
D.~Nishida\altaffilmark{1}
K.~Nishijima\altaffilmark{10}
M.~Ohishi\altaffilmark{3}
Y.~Sakamoto\altaffilmark{10}
A.~Seki\altaffilmark{10}
V.~Stamatescu\altaffilmark{5}
T.~Suzuki\altaffilmark{2}
D.~L.~Swaby\altaffilmark{5}
G.~Thornton\altaffilmark{5}
F.~Tokanai\altaffilmark{7}
K.~Tsuchiya\altaffilmark{1}
S.~Watanabe\altaffilmark{1}
Y.~Yamada\altaffilmark{11}
E.~Yamazaki\altaffilmark{10}
S.~Yanagita\altaffilmark{2}
T.~Yoshikoshi\altaffilmark{3}
Y.~Yukawa\altaffilmark{3}
}
\email{nakamori@cr.scphys.kyoto-u.ac.jp}

\altaffiltext{1}{ Department of Physics, Kyoto University, Sakyo-ku, Kyoto 606-8502, Japan} 
\altaffiltext{2}{ Faculty of Science, Ibaraki University, Mito, Ibaraki 310-8512, Japan} 
\altaffiltext{3}{ Institute for Cosmic Ray Research, University of Tokyo, Kashiwa, Chiba 277-8582, Japan} 
\altaffiltext{4}{ Research School of Astronomy and Astrophysics, Australian National University, ACT 2611, Australia} 
\altaffiltext{5}{ School of Chemistry and Physics, University of Adelaide, SA 5005, Australia} 
\altaffiltext{6}{ CSIRO Australia Telescope National Facility, Narrabri, NSW 2390, Australia} 
\altaffiltext{7}{ Department of Physics, Yamagata University, Yamagata, Yamagata 990-8560, Japan} 
\altaffiltext{8}{ Ibaraki Prefectural University of Health Sciences, Ami, Ibaraki 300-0394, Japan} 
\altaffiltext{9}{ Faculty of Management Information, Yamanashi Gakuin University, Kofu, Yamanashi 400-8575, Japan} 
\altaffiltext{10}{ Department of Physics, Tokai University, Hiratsuka, Kanagawa 259-1292, Japan} 
\altaffiltext{11}{ Department of Physics, Konan University, Kobe, Hyogo 658-8501, Japan} 
\altaffiltext{12}{ Department of Physical Science, Hiroshima University, Higashi-Hiroshima, Hiroshima 739-8526, Japan} 
\altaffiltext{13}{ Solar-Terrestrial Environment Laboratory,  Nagoya University, Nagoya, Aichi 464-8602, Japan} 
\altaffiltext{14}{ National Astronomical Observatory of Japan, Mitaka, Tokyo 181-8588, Japan} 
\altaffiltext{15}{ School of Allied Health Sciences, Kitasato University, Sagamihara, Kanagawa 228-8555, Japan} 
\begin{abstract}
We have observed the supernova remnant MSH\,15$-$52 (G320.4$-$1.2),
which contains the gamma-ray pulsar PSR B1509$-$58,
using the CANGAROO-III imaging atmospheric Cherenkov telescope array
from April to June in 2006.
We detected gamma rays above 810\,GeV
at the 7 sigma level during a total effective exposure of 48.4~hours.
We obtained a differential gamma-ray flux at 2.35\,TeV of 
$(7.9\pm 1.5_{\rm stat}\pm 1.7_{\rm sys}) \times 10^{-13}$\,cm$^{-2}$\,s$^{-1}$\,TeV$^{-1}$
with a photon index of $2.21 \pm 0.39_{\rm stat} \pm 0.40_{\rm sys}$,
which is compatible with that of the H.E.S.S.\ observation in 2004.
The morphology shows extended emission
compared to our Point Spread Function.
We consider the plausible origin of the high energy emission
based on a multi-wavelength spectral analysis and energetics arguments.
\end{abstract}

\section{Introduction}
Recent imaging atmospheric Cherenkov telescopes (IACTs)
have achieved remarkably high sensitivity
in the very high energy gamma-ray band.
This is well illustrated by 
the Galactic plane survey carried out by the H.E.S.S.\ collaboration
\citep{aha06,hop07}.
Many of the discovered TeV sources are
associated with pulsar wind nebulae (PWNe) \citep[e.g.,][]{gae06},
which are now established as the most populous category among Galactic TeV
sources: 18 PWNe have been found so far \citep{hin07} and
a portion of 21 unidentified galactic TeV sources could be PWNe as well
\citep{aha07b,fun07b}.

The Crab nebula, the prototype PWN, 
has been observed in every accessible wave band. 
The nebula contains the Crab pulsar,
which has the highest spindown energy loss
of known gamma-ray pulsars \citep{tho03}.
However, the radiation mechanism
remains controversial for the Crab nebula and for other PWNe:
it has been argued that the Crab nebula's 
broadband spectral energy distribution (SED) can be
well explained by synchrotron emission in an 
average magnetic field of $\sim 0.1-0.3$\,mG
for radio to soft gamma-ray bands, and by inverse Compton (IC) scattering of
synchrotron, IR, millimeter and cosmic microwave background (CMB)
photons for the hard gamma-ray band up to 100\,TeV
 \citep{ahat96,ato96,aha97,aha98,aha04}.
On the other hand, it was claimed that
gamma rays from the decay of neutral pions, produced by hadrons,
contribute significantly at $\sim 10$\,TeV for the Crab nebula \citep{bed03}.
Consideration has also been given to
whether the energy source of TeV gamma-ray emission
from an SNR containing a pulsar (composite SNR) 
is the pulsar's spindown
energy or related to the supernova explosion \citep[e.g.][]{fun07b},
and the efficiency of energy conversion to the particle acceleration
for such models.

PSR B1509$-$58 has the third highest spindown energy loss
after the Crab pulsar \citep{tho03} and PSR J1833-1034 \citep{cam06}
in the Galaxy, and its nebula
has also been well studied across the electromagnetic spectrum.
TeV gamma-ray observations, combined with those at other energy bands,
provide additional information which 
may lead to solutions for the above problems and a unified comprehension
of pulsar and nebula systems.
PSR B1509$-$58 was detected
in the radio supernova remnant MSH\,15$-$52 (G320.4$-$1.2) \citep{cas81},
initially as a 150-ms X-ray pulsar with the {\it Einstein} 
satellite \citep{sew82},
which was confirmed by later X-ray/soft gamma-ray observations
\citep{kaw93,mat94,gun94,sai97,mar97,cus01,for06}.
Subsequently its pulse period was detected at radio frequencies \citep{man82},
and at soft gamma-ray energies with COMPTEL \citep{kui99}.
The pulsar was detected above 30\,MeV by EGRET at the 4.4$\sigma$ level
with some suggestive, if not statistically compelling, evidence 
of modulation at the pulsar period \citep{kui99}.
Although optical \citep{she98} and near-IR \citep{kap06} searches
found possible pulsar counterparts,
the pulse period was not detected.
As one of the most energetic young pulsars, PSR B1509$-$58,
has been particularly well studied at radio wavelengths.
A detailed timing analysis yielded
a period derivative of $\dot{P} = 1.5\times 10^{-12}$,
and a high spin-down luminosity of
$\dot{E} = 1.8\times 10^{37}I_{45}$\,ergs\,s$^{-1}$,
where $I_{45}$ is the moment of inertia in units of $10^{45}$g cm$^{-2}$.
A braking index of $n = 2.84$ was measured, corresponding to
an age, assuming an initial period of $P_0 \ll P$, of
$\tau = (P/(n-1)\dot{P})[1-(P_0/P)^{n-1}]
\sim(P/(n-1)\dot{P}) \sim 1700$\,yr.
A large dipole surface magnetic field of 
$B=1.5\times10^{13}$\,G was also inferred \citep{kas94,liv05}.
However, in contrast to other young pulsars, no pulsar glitch has been 
observed to date.
Initially, the age of the SNR was estimated to be
$6-20$\,kyr \citep{sew83} or  $\sim 10$\,kyr \citep{van84}
prompting debate about the disagreement with the pulsar's age.
\citet{bla88} and \citet{gva01}, for instance, suggested
the pulsar age was actually $\ge 20$\,kyr.
The latter was based on the assumption
that pulsar's braking torque was enhanced
by the interaction between the pulsar's magnetosphere and
circumstellar dense clumps.

The radio morphology of MSH\,15$-$52
consists of southeast and northwest shells.
The latter, $\sim 10'$ from the pulsar,
spatially coincides with the H$\alpha$ nebula RCW89 \citep{rod60},
and \citet{gae99} concluded that MSH\,15$-$52,
PSR B1509$-$58 and RCW89 were associated systems.
The distance, derived from an H\,I absorption measurement,
is $5.2 \pm 1.4$\,kpc, 
consistent with the value of $5.9 \pm 0.6$\,kpc
determined from the pulsar dispersion measure \citep{tay93}.
We adopt $d=5.2$\,kpc throughout this paper.

Symmetric jets,
similar to the Crab pulsar \citep{bri85} and the Vela pulsar \citep{hel01},
were observed by {\it ASCA} \citep{tam96},
{\it ROSAT} \citep{tru96} and {\it Chandra} \citep{gae02}.
The jet directed towards the northwest was
observed to terminate at RCW89.
Precise {\it Chandra} observations revealed
the sequential heating of the knots in RCW89
by the pulsar jet \citep{yat05}.
The high resolution {\it Chandra} image also showed
the arc structure where the pulsar wind may be terminated,
and the diffuse pulsar wind nebula \citep{gae02},
which emitted non-thermal synchrotron radiation.
However the PWN has not been observed
at other wavelengths such as IR or optical.
Although IRAS found an infrared source, {\it IRAS} 15099$-$5856,
spatially coincident with PSR B1509$-$58,
the IR emission was thermal and
therefore not related to the PWN \citep{are91}.
A faint radio structure was detected \citep{gae99} but
its flux density was obtained only
within a large error \citep{gae02}.
A larger extent than that of X-ray was expected
due to the difference of the cooling lifetime so that
it was thought to be partially hidden by the bright RCW89.

The {\it Ginga} LAC discovered single power-law emission up to 20\,keV
with a photon index of $\sim2$, indicating synchrotron emission
and the existence of accelerated electrons \citep{asa90}.
The synchrotron nebula spectrum was also measured 
by {\it EINSTEIN} \citep{sew83}, {\it EXOSAT} \citep{tru90} and
{\it RXTE} \citep{mar97}.
{\it BeppoSAX} detected nonthermal emission from 1\,keV up to 200\,keV
with a photon index of $\Gamma = 2.08\pm 0.01$ \citep{min01},
while recent observations with {\it INTEGRAL} IBIS
found a possible ($2.9\sigma$) spectral cut off at $\sim 160$\,keV 
\citep{for06}.
Since high energy electrons were demonstrated to exist,
very high energy gamma-ray emission was predicted
via IC scattering with CMB photons \citep{du95,har96}.
CANGAROO-I also suggested a possible VHE gamma-ray detection
of $\sim 10 \%$ of the Crab flux above 1.9\,TeV,
assuming a spectral photon index of 2.5 \citep{c1}.
H.E.S.S.\ subsequently reported extended VHE gamma-ray emission
along with the pulsar jet.
Their TeV morphology showed a good coincidence with X-ray images,
indicating that the TeV gamma-rays originate from
the inverse Compton scattering of relativistic electrons.
It was pointed out that
IC scattering of the CMB could not account for
most of the TeV gamma-ray flux with an assumed magnetic field of 17~$\mu$G,
which indicated the contribution of IR photons
as seed photons for the IC process \citep{aha05,khe05}.
The necessity for target photons in addition to the CMB
has recently been suggested for other Galactic sources also
\citep{khe05,hin07a}.
Here we report the results of TeV gamma-ray observations with the
CANGAROO-III telescopes
and consider the origin of the TeV gamma-ray emission based on a 
discussion of the energetics.

%
%
%

\section{CANGAROO-III Observations}

CANGAROO-III is an array of four IACTs, 
located at Woomera, 
South Australia (136$^{\circ}47'$E, $31^{\circ}06'$S, 160m a.s.l.). 
Each telescope has a 10\,m diameter reflector 
which consists of 114 segmented FRP spherical mirrors
mounted on a parabolic frame \citep{kaw01}. 
The telescopes are situated at the corners of a diamond 
with $\sim$100\,m sides \citep{eno02}. 
The oldest telescope, T1, which was the CANGAROO-II telescope, 
was not used due to its smaller FOV and higher energy threshold. 
The imaging camera systems on the other three telescopes (T2, T3 and T4) 
are identical, with 427 PMTs and a FOV of 4.0$^{\circ}$ \citep{kab03}.
The PMT signals were recorded by charge ADCs and multi-hit TDCs \citep{kub03}.
The observations were made from April to June in 2006. 
The tracking positions were offset 
by $\pm 0.5^{\circ}$ from PSR B1509$-$58 
in declination or in right ascension,
and changed every twenty minutes,
in order to
suppress position-dependent effects on the camera 
due to bright (4.1 and 4.5 magnitude) stars.
To trigger data recording, an individual telescope was required to
have more than four pixels registering over 7.6 photoelectrons within 100\,nsec
(local trigger),
with a global trigger system then determining the coincidence 
of any two of the three telescopes \citep{c3trig}. 
We rejected data taken in bad weather conditions 
in which the shower event rate was less than 5\,Hz
or at zenith angles larger than $35^{\circ}$.
Finally, the selected data were taken 
at a mean zenith angle of $30.1^{\circ}$ 
with a corresponding Point Spread Function (PSF)
of $0.23^{\circ}$ (68\% containment radius).
A typical trigger rate of 3-fold coincidence is 12\,Hz.
The effective exposure time amounts to 48.4\,hours.
%
%
%
%
%
%
%
\section{Data reduction and Analysis}

The basic analysis procedures are described
in detail in \citet{eno06a} and \citet{kab07}.
Using calibration data taken daily with LEDs, the
recorded charges of each pixel in the camera 
were converted to photoelectrons.
At this step we found 8 bad pixels out of $427\times3$
due to their higher or lower ADC conversion factors
in these observations.
These bad pixels were removed from this analysis period,
which was also reflected in the Monte Carlo simulations.
Then every shower image was cleaned through
the following CANGAROO-III standard criteria.
Only pixels which received $\ge$5.0 photoelectrons were used as ``hit pixels''.
Then five or more adjacent hit pixels, with arrival times
within 30\,nsec from the average hit timing of all pixels,
were recognized as a shower cluster.

We carefully studied the effect of the bright stars
by monitoring the mean ADC counts of the pixels for which
the stars entered the FOV.
When the stars were within the pixel's FOV, the PMT 
hit rate increased significantly.
At such times the ``hits'' for that pixel were dominated by the 
starlight triggers,
which were dimmer than ``hits'' by air shower Cherenkov photons.
The average ADC counts 
of the affected pixels were clearly reduced during such times.
After the image cleaning procedure above,
the averaged ADC counts became stable 
within the usual values over the whole run,  
indicating the effects of the stars were completely removed.

Before calculating image moments 
--- the ``Hillas parameters" \citep{hil85} ---
we applied the ``edge cut'' described in \citet{kab07}.
If the brightest 15 pixels in the image  
(or all pixels for images with less than 15 hits) 
were not located in the outer edge layer of the camera,
we retained the event.
The orientation angles were determined 
by minimizing the sum of squared {\it{width}}s 
with a constraint given by the {\it{distance}}
predicted by the Monte Carlo simulations. 

Then we applied the Fisher Discriminant method \citep{fis36,eno06a} 
with a multi-parameter set of $\vec{P} = (W_2,W_3,W_4,L_2,L_3,L_4)$, 
where $W$ and $L$ are the energy corrected {\it{width}} and {\it{length}}, 
and suffixes identify the telescope.
The Fisher Discriminant (FD) is defined 
as $FD \equiv \vec{\alpha}\cdot \vec{P}$ 
where $\vec{\alpha}$ is a set of coefficients mathematically determined 
to maximize the separation between FD for gamma-rays and hadrons.

\begin{figure}[t]
\epsscale{}
\plotone{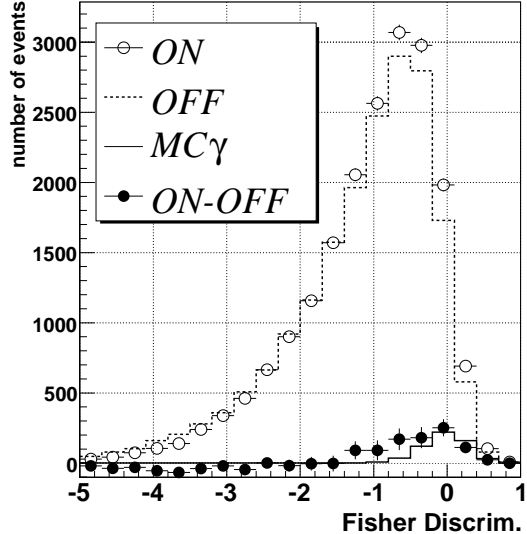}
\caption{FD distribution. 
  The open circles show the FD obtained 
  from the ON source region, $\theta ^2<0.1$\, deg$^2$. 
  The broken and solid histogram are 
  the background and gamma-ray component 
  estimated by the fit procedure described in the text. 
  The filled circles are the subtraction of the background from the ON source region.}
\label{fd}
\end{figure}

For the background study we selected a ring region around the target, 
$0.2$\,deg$^2$$\le \theta ^2 \le0.5 $\,deg$^2$, 
and obtained FD distributions 
for background, $F_{b}$, and Monte Carlo gamma-rays, $F_g$. 
Finally we could fit the FD distributions of the events from the target
with a linear combination of these two components.
The observed FD distributions, $F$, should be represented as
$F = \alpha F_g + (1-\alpha)F_b$
where $\alpha$ is the ratio of gamma-ray events to total events.
Here only $\alpha$ is optimized and
the obtained FD distributions are shown in Fig.\ \ref{fd}.
This analysis method was checked 
by an analysis of the Crab nebula data taken in December 2005.

The reflectivities of each telescope, which are used in the simulations,
are monitored every month 
by a muon ring analysis of calibration run 
individually taken by each telescope. 
We obtained relative light collecting efficiencies 
with respect to the original mirror production time
of  0.61, 0.64, and 0.67 for T2, T3 and T4, respectively.  

%
%
%
%
\section{Results}

\begin{figure}[tb]
\epsscale{}
\plotone{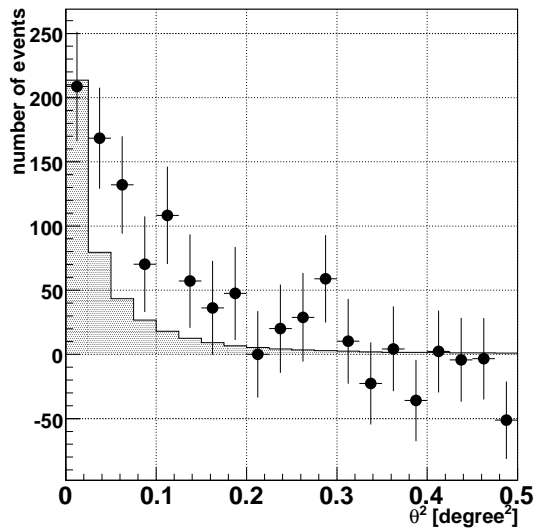}
\caption{The $\theta ^2$ plot. 
  Here, 0 deg corresponds to the best-fit position of the centroid of the emission
  obtained by this work (see text).
  The hatched histogram represents our PSF derived from the Monte-Carlo simulation.
} 
\label{theta}
\end{figure}
\begin{figure}[t]
\epsscale{}
\plotone{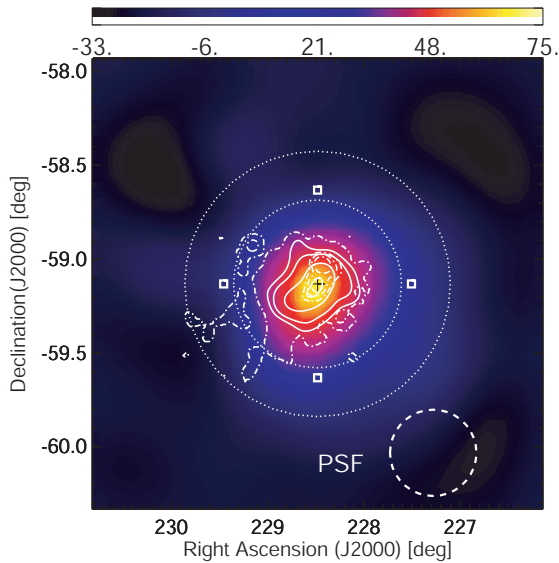}
\caption{Morphology of gamma-ray like events,
  smoothed with a gaussian of $\sigma=0.23^\circ$.
  Our PSF is also shown by a dashed circle(68\% containment radius).
  The squares and the cross represent 
  tracking positions and the pulsar position, respectively. 
  Solid contours show VHE gamma-ray emission 
  as seen by H.E.S.S.\ and dotted contours by {\it ROSAT} 0.6-2.1\,keV \citep{tru96}. 
  The region between thin dotted circles are used for the background study.}
\label{map}
\end{figure}
%
The obtained $\theta ^2$ plot is shown in Fig.\,\ref{theta}
with the PSF of our telescopes.
Above 810\,GeV we detected 427$\pm$63 excess events
under the assumption the TeV source
was a point source ($\theta ^2<0.06$\,deg$^2$) 
and 582$\pm$77 events (7.6$\sigma$)
within $\theta ^2 <0.1$\,deg$^2$, which corresponds to the size of the SNR.
The TeV gamma-ray emission is extended, and 
the morphology of gamma-ray--like events,
smoothed by a gaussian with $\sigma=0.23$~deg,
is shown in Fig.\,\ref{map}.
The number of events were individually estimated
by the FD-fitting method 
in each $0.2^\circ \times 0.2^\circ$ sky bin.
When we evaluate the outer regions ($\theta ^2>0.6$\,deg$^2$), 
we must consider the gradual deformations 
of the FD distributions at larger angular distances from the target.
Therefore we selected a ring with radii $0.2^{\circ} < r < 0.4^{\circ}$
centered on the evaluated region as the background.
For the inner regions, $\theta ^2 <0.6$\,deg$^2$, 
the same background as used in Fig.~\ref{theta} was adopted.
The intrinsic extent of the TeV gamma-ray emission
was estimated by the 2D Gaussian fit 
on our unsmoothed excess map.
The inclination of the major axis is
$61.3 \pm 1.9^\circ$, 
measured to the north from west, 
which is compatible with the jet direction determined by {\it Chandra},
H.E.S.S. and {\it INTEGRAL}.
The intrinsic source sizes along the major and minor axes were 
calculated to be $0.07\pm 0.07^\circ$ and $0.21\pm 0.08^\circ$, respectively. 
The center of gravity of the TeV gamma-ray emission
is at (R.A., Dec.)=$(228^\circ .486, -59^\circ .235)$,
which corresponds to the offset from the pulsar in (R.A., Dec.) =
$(0^\circ .0030\pm 0^\circ .0076,\, 0^\circ .10\pm 0^\circ .012)$.
No significant offset from the pulsar is then observed given our PSF,
while the H.E.S.S. measurement showed the offset
of $(0^\circ .048, 0^\circ .022)$ in (R.A., Dec.) at the $3 \sigma$ level. 
%
%
%
%
\begin{figure}[t]
\epsscale{}
\plotone{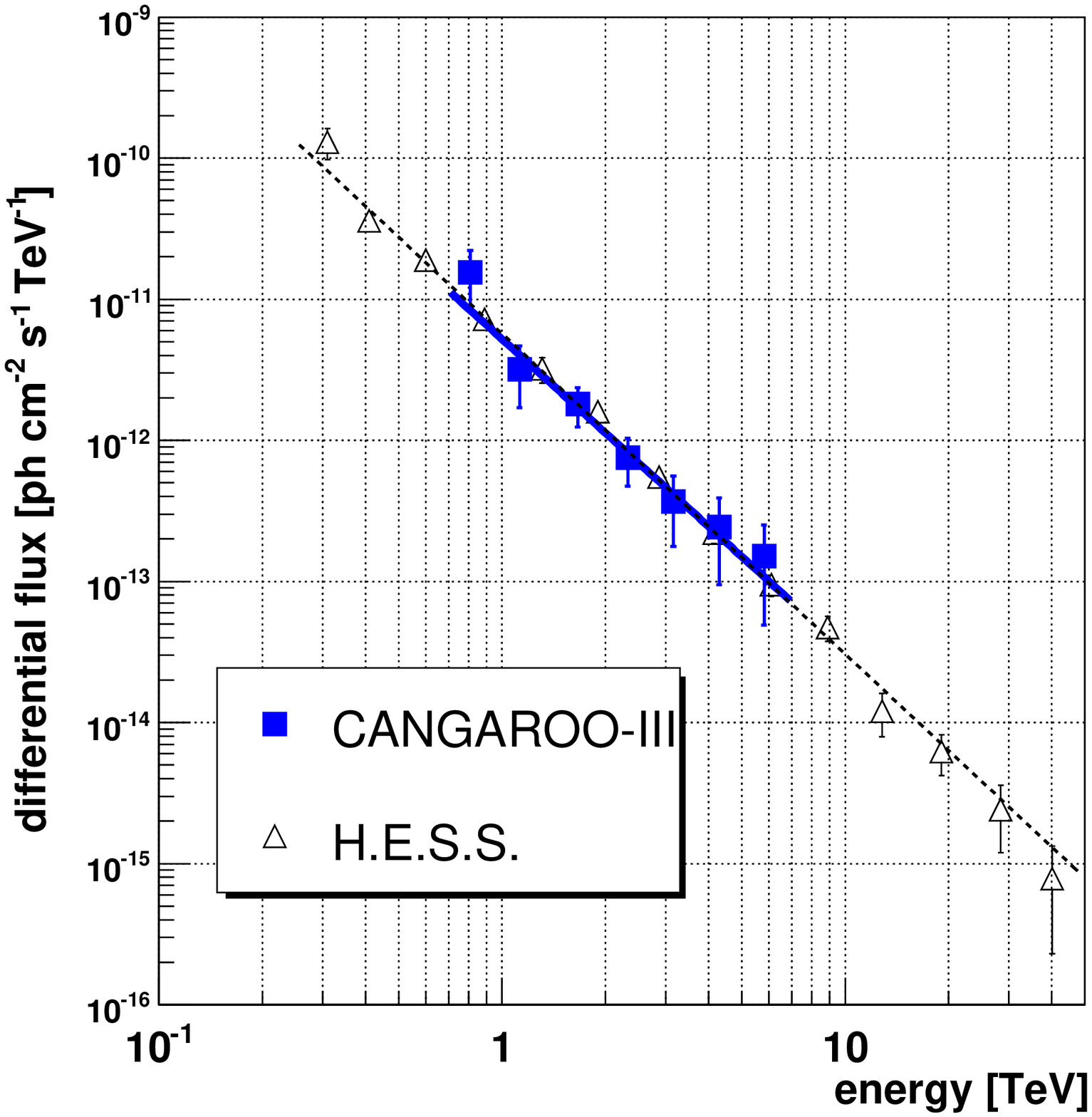}
\caption{Differential flux of the whole nebula.
  Squares and triangles show the CANGAROO-III 
  and the H.E.S.S.\ data points, respectively.
  The best fit power-laws are also shown
  by the solid and dashed line from 
  this work and from H.E.S.S., respectively.
}
\label{flux} 
\end{figure}
%
%
%
Fig.\,\ref{flux} represents a reconstructed VHE gamma-ray spectrum
compatible with a single power-law:
the value at 2.35\,TeV is
$(7.9\pm 1.5_{\rm stat}\pm 1.7_{\rm sys}) 
\times 10^{-13}$ cm$^{-2}$s$^{-1}$TeV$^{-1}$ 
with a photon index of 
$2.21 \pm 0.39_{\rm stat} \pm 0.40_{\rm sys}$.
The flux was measured within $\theta ^2 < 0.1$\,deg$^2$.
The relevant systematic errors are due to the atmospheric transparency,
NSB fluctuations, uniformity of camera pixels, and
light collecting efficiencies.
In addition, the signal integrating region was changed 
from $\theta ^2 <0.1$~deg$^2$ to 0.2~deg$^2$ in Fig.\,\ref{theta},
which was included in the systematic error.
The TeV gamma-ray extent and flux obtained by CANGAROO-III
are consistent with those obtained with H.E.S.S.
Our result indicates the TeV gamma-ray emission does not vary
significantly between the H.E.S.S.\ observations in 2004 and ours in 2006,
which is also consistent with the steady X-ray emission 
from the diffuse PWN over several decades \citep{del06}.
\citet{c1} reported a marginal detection (4.1$\sigma$) above 1.9\,TeV 
of $(2.9 \pm 0.7_{\rm stat})\times 10^{-12}$\,ergs\,cm$^{2}$\,s$^{-1}$.
For the sake of comparison,
we integrated the differential flux of this work over the same energy band
with a photon index of 2.5 as inferred by \citet{c1}, and obtained
$(1.6\pm 0.4_{\rm stat})\times 10^{-12}$ ergs\,cm$^{-2}$\,s$^{-1}$, 
marginally outside the 1$\sigma$ error range of the CANGAROO-I result.
%
%
%
%
%
%
\section{Discussion}
We discuss the possible origin of the TeV gamma-ray emission 
from the PWN for both proposed cases of its age: 
$\tau =1700$\,yr and 20\,kyr \citep{bla88,gva01}. 
\subsection{Upper limit on the global energetics}

In general, two alternatives of the energy source of a composite SNR
could be considered,
namely the pulsar's spindown energy and
the supernova explosion (or the SNR).
In the latter case,
the energy could be as high as $\sim 10^{51}$\,ergs,
or even up to a few times $10^{52}$\,ergs for some SNe\,Ib/c and
SNe\,IIn \citep{nom01}.
However, in the case of MSH 15$-$52,
the extent of the intrinsic TeV gamma-ray morphology
shows a good coincidence
not with the shell of MSH\,15$-$52 but with the jet and the PWN.
Therefore we rejected the scenario 
in which the SNR is at the origin of the high energy particles.
Hereafter we assume the pulsar's spindown energy 
is the global energy source.

The time evolution of the pulsar period $P(t)$ is described as
\begin{equation}
P(t)=P_0\Bigl(1+ \frac{t}{\tau _0}\Bigr)^{\frac{1}{n-1}},
\end{equation}
where $\tau _0$ is a parameter called the initial spindown time scale,
assuming both $n$ and $k$ in the braking equation 
$\dot{\Omega} = -k\Omega^n$ are constant \citep[e.g.,][]{gae06}.
The pulsar's spindown energy $\dot{E}(t)$ is then calculated as
\begin{equation}
  \dot{E}(t) = \dot{E_0}\Bigl(1+\frac{t}{\tau _0}\Bigr)^{-\alpha}, \>\>  \alpha = \frac{n+1}{n-1}
\end{equation}
where $\dot{E_0}$ is the initial spindown luminosity.
By integrating this, we obtain the total energy 
which the pulsar has lost over its age, $\tau$,
\begin{equation}
E_{\rm tot} =\int _0 ^{\tau}\dot{E}(t)dt= \frac{\dot{E_0}\tau _0}{1-\alpha}\Bigl[\Bigl(1+\frac{\tau}{\tau _0}\Bigr)^{1-\alpha} -1\Bigr].
\label{etot}
\end{equation}
This formula provides an upper limit to the energy supplied by the
pulsar to the PWN as a function of the unknown parameter $\tau _0$.
For the sake of simplicity, energy loss through adiabatic expansion
was neglected here.  
The  dependence of $E_{\rm tot}$ on $\tau _0$ is listed in 
Table~\ref{tau0}, where the pulsar's moment of
inertia was assumed to be $10^{45}$\,g\,cm$^{-2}$.  
The value of $\tau _0$ also determines
the initial spin period $P_0$, using the current period $P(\tau)$,
as listed in Table~\ref{tau0}. 
For comparison, the $E_{\rm tot}$ and $P_0$ for the Crab pulsar were also
listed in Table~\ref{tau0}, 
with $P = 33$\,msec, $\dot{P}=4.2\times 10^{-13}$ \citep{tay93}, 
$\tau = 950$\,yr, $\dot{E}(\tau)=5\times 10^{38}$\,ergs\,s$^{-1}$
 and $n=2.5$ \citep{lyn88}.  
The $E_{\rm tot}$ of
the Crab is about 10 times larger than that of PSR B1509$-$58 for the
same $\tau _0$.

For $\tau = 1700$\,yr, we can calculate $\tau _0$ as
\begin{equation}
  \tau _0 = \frac{P(\tau)}{(n-1)\dot{P}(\tau)} -\tau \sim 30 {\rm yrs}.
\label{eq4}
\end{equation}
However, in the $\tau = 20$\,kyr case,
these formulae are not applicable
due to the possibility of a time-dependence for $k$ \citep{bla88}.
It is hard to estimate an accurate $E_{\rm tot}$, however,
a larger $E_{\rm tot}$ than that estimated for $\tau =1700$\,yrs
would be expected.

\begin{deluxetable}{lccc} 
\tablewidth{0pt}
\tablecaption{$\tau _0$ dependence of parameters.
\label{tau0}}
\tablehead{
  \colhead{PSR}&
  \colhead{$\tau _0$ [yr]}&
  \colhead{$E_{tot}$ [ergs]}&
  \colhead{$P_0$ [msec]}
}
\startdata
B1509$-$58& 30 & $7.5\times 10^{49}$ & 16\\
($\tau$=1700 yr) &100 & $2.1\times 10^{49}$ & 31\\
              &300 & $7.2\times 10^{48}$ & 53\\
              &500 & $4.6\times 10^{48}$ & 67\\
              &700 & $3.6\times 10^{48}$ & 77\\
              &1000 & $2.8\times 10^{48}$ & 87\\
Crab          & 30\tablenotemark{a}  & $1.1\times 10^{51}$ & 3.3\\
              & 700\tablenotemark{b} & $3.7\times 10^{49}$ & 19
\enddata
\tablenotetext{a}{\citet{ato99} argued that
the observed radio spectrum suggested $\tau _0 = 30$\ yr.}
\tablenotetext{b}{derived from eq.~(\ref{eq4}). }
\end{deluxetable}

%
%
%
\subsection{Hadronic scenario}

\begin{figure}[tb]
\epsscale{}
\plotone{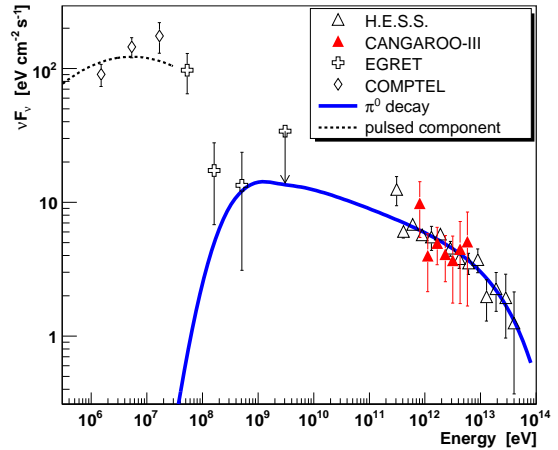}
\caption{Spectral energy distribution and the model curve of 
  neutral pion decay. 
  The solid line represents the best fit model curve.
  COMPTEL and EGRET flux points were omitted from the fit 
  since they contained pulsed emission from the pulsar. 
  A dashed line cited from \citet{cus01} 
  represents a model fit curve 
  of the pulsed emission obtained by{\it BeppoSAX} and COMPTEL.
}
\label{pion}
\end{figure}

Hadronic gamma-ray production in PWNe has been suggested
for the Crab nebula \citep{bed03,ama03}
 and the Vela~X region \citep{bed07,hor06,hor07}.
First we examine a neutral-pion decay model 
for the origin of the TeV gamma-ray emission.
Fig.~\ref{pion} shows the spectral energy distribution (SED).
We assumed the population of accelerated protons
to be expressed by a single power-law with an exponential cutoff,
that is, $dN_p/dE_p \propto E_p^{-\gamma _p}\exp (-E_p/E_{\rm max})$.
We used only our data and the H.E.S.S.\ data \citep{aha05} for the model fit
since the COMPTEL and EGRET data were the sum of
pulsed and unpulsed emission \citep{kui99}
which may contain emission from the pulsar.
The contribution of the pulsed emission is indicated by the dashed line in
Fig.~\ref{pion},
derived from a fit to the {\it BeppoSAX} and COMPTEL data of
the pulsed emission \citep{cus01} (see also Fig.~\ref{pulse}).
Since these fluxes were apparently dominated by the pulsed emission,
we didn't use them in the model fit.
\citet{aha07} put upper limits to the TeV gamma-ray pulsed emission
of approximately one order magnitude below the total flux,
and this component was excluded from the fit.
The best-fit curve is shown in Fig.~\ref{pion} by the solid line.
The power-law index and the cutoff energy were obtained
to be $\gamma _p=2.16\pm0.05$ and $E_{\rm max}=530\pm399$\,TeV, respectively.
The total energy of high energy protons above 1\,GeV, $W_p$, was
calculated to be $W_p = 3.2\times 10^{51}(n/1 {\rm cm^{-3}})^{-1}$\,ergs,
which means that the pulsar is not able to produce the TeV gamma-ray emission
simply with an interstellar matter (ISM) density of 1\,cm$^{-3}$.
\citet{dub02} reported a denser ISM distribution
of $n \sim 10$~cm$^{-3}$, derived from the H\,I observations
with ATCA, which is indeed valid only in the northwest radio limb.
If such a high density was uniformly applicable,
the total energy would be reduced to $W_p=3.2 \times 10^{50}$\,ergs.
\citet{dub02}, however,  also mentioned that
the southeast radio limb of the SNR showed
$n \sim 0.4$~cm$^{-3}$,
which would yield to a higher $W_p$.
In any case, it is revealed that the 
spin-down energy can not drive the acceleration of protons
during its characteristic age,
even if the uncertainty in the distance (see \S\,1) is taken
into account.
Therefore the hadronic gamma-ray production
originated by the pulsar spin-down energy was strongly unlikely. 
Note that bremsstrahlung emission is not dominant in the TeV gamma-ray band
under these ambient densities.

%
%
%
%
%
\subsection{Leptonic scenario}

Secondly, we discuss the leptonic scenario, and 
here we used a simple one-zone IC model 
to reproduce the multi-wavelength spectra.

\begin{figure}[t]
\epsscale{}
 \plotone{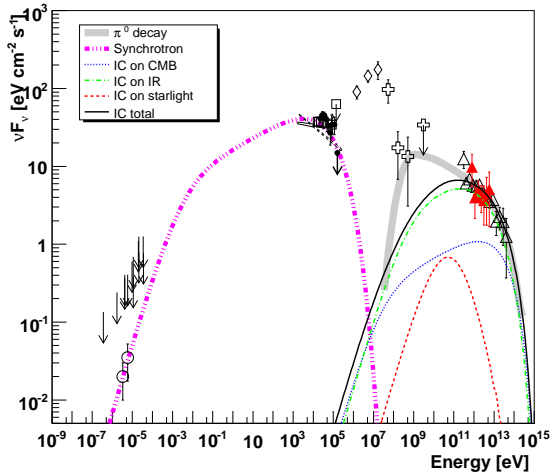}
\caption{SED and leptonic model curves. 
  The references of data are given in Table~\ref{seddata}.
  Assumed electron spectrum was a broken power-law
  and the derived model parameters are listed in Table~\ref{summary}.
  The IC spectra on each of CMB, IR and star light 
  are represented by dotted, dot-dashed and dashed curves, respectively. 
  The pion decay model curve in Fig.~\ref{pion} is 
  shown by the thick curve for comparison.}
\label{sync1}
\end{figure}

\begin{figure}[t]
\epsscale{}
\plotone{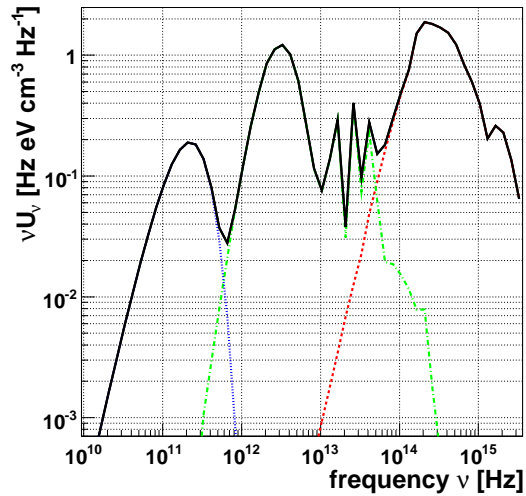}
\caption{
  Interstellar radiation field from the GALPROP package (v50p)
  at $(R, z)=(5.6, -0.11)$~kpc.
  From lower frequencies, CMB (dotted),
  IR emission from interstellar dust (dot-dashed),
  and optical photons from stars (dashed) 
  are shown.
  The solid line represents the sum of the three components. 
 }
\label{isrf}
\end{figure}

\noindent
{\bf Case I}: The multi-band SED is plotted in Fig.~\ref{sync1}. 
The data points derived from this work are 
represented by filled triangles
and references to others are summarized in Table~\ref{seddata}.
Since {\it BeppoSAX}/PDS data in \citet{min01} 
should be corrected by the intercalibration factor of
about 20\% between MECS and PDS,
here we plotted the corrected PDS data
(Mineo 2007, private communication).
The highest energy bin of the PDS represents a 1\,$\sigma$ upper limit.
IBIS measurement reported extended emission
with a possible spectral cutoff at about 160\,keV,
at the 2.9\,$\sigma$ confidence level 
from the extrapolation of the power-law spectrum.
Then we present an upper limit in Fig.~\ref{sync1}
derived from multiplying the 1\,$\sigma$ upper limit by a factor of 3.
Note that, though the data analysis of the coded-mask instrument of IBIS
is ideally designed for point-like sources,
\citet{for06} have extracted the spectrum of the observed extended emission
following a suited method developed by \citet{ren06}.
Arrows in the radio band show the whole emission
from MSH\,15$-$52 or RCW\,89 listed in \cite{du95},
which we treated as upper limits for the faint PWN.
As well as in the case of the hadronic scenario,
the COMPTEL and EGRET fluxes were not used as they
contain a component of pulsed emission \citep{kui99}.
The CMB field density on its own could not account for the TeV gamma-ray 
flux, and hence
we added IR and optical (starlight) photon fields, following \cite{aha05}. 
Here we used an interstellar radiation field (ISRF)
derived from the latest (v50p) GALPROP package 
\citep{por05,gal06}.
The ISRF is given for three components
(CMB, IR from dust, and optical starlight)
as a function of a distance from the Galactic center $R$ (in kpc)
and 
distance from the Galactic plane
$z$ (in kpc).
We extracted the spectra at $(R,z)=(5.6, -0.11)$,
which is shown in Fig.~\ref{isrf},
and the nominal value is 1.4 eV/cc both for IR and optical light.
The ISRF field densities did not change drastically
when we considered the ambiguity in the distance, 
within 1.4-0.89 eV/cc and 1.4-0.90 eV/cc for IR and optical, respectively.
Note that these values extracted from GALPLOP do not represent the local densities.

%
%
The best-fit curve is shown in Fig.~\ref{sync1}
with the assumption of a broken power-law electron spectrum;
\begin{equation}
\frac{dN_e}{dE_e} \propto \frac{(E _e/E _{\rm br})
^{-\gamma_1}}{1+(E_e/E_{\rm br})^{(\gamma _2-\gamma _1)}}\exp{(-E
_e/E _{\rm max})}
\label{brokenpl}
\end{equation}
where $E_e$ is the electron energy,
$\gamma_1$ is the spectral index of the injected electrons,
$\gamma_2$ is that of the cooled electrons, and
$E_{\rm br}$ and $E_{\rm max}$ are the break 
and maximum electron energies, respectively.
The derived parameters are listed in Table~\ref{summary}.

\begin{figure}[t]
\epsscale{}
\plotone{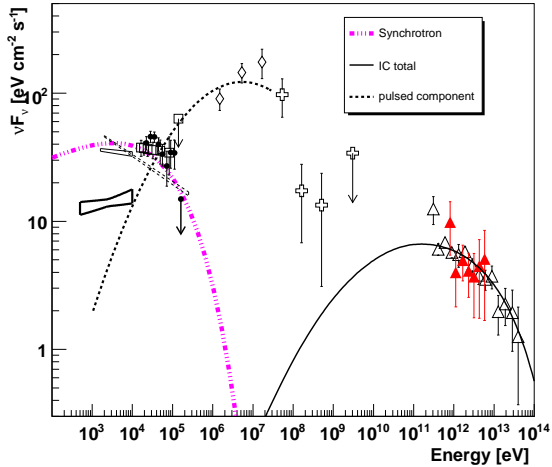}
\caption{
  {\it Chandra} flux is represented by a region closed with thick lines.
  Other data and the synchrotron and IC model curve in Fig.~\ref{sync1}
  are also shown.
}
\label{pulse}
\end{figure}

Fig.~\ref{pulse} shows a close-up of the SED and the derived model curves.
According to the IBIS measurement,
the unpulsed emission is dominated by the PWN \citep{for06},
which is also supported by the fact that
the timing analysis showed unpulsed emission from the pulsar is
indeed several factors less than the pulsed component\citep{cus01,for06}.
The model curve well reproduced the hard X-ray spectra.
The region closed with thick solid lines in Fig.~\ref{pulse} 
represents {\it Chandra} measurement excluding the pulsar,
which corresponds to the sum of 
the ``diffuse PWN'', ``jet'' and ``outer arc'' fluxes
listed in \citet{del06}.
The FOV of the {\it Chandra} measurement was 
smaller than the MECS signal region,
which may cause the apparent discrepancy in the soft X-ray band.
Therefore we didn't take care of the data for the fit.

The obtained magnetic field of 17~$\mu$G was consistent with \citet{aha05},
and higher than previous indications \citep[][and references therein]{gae02}. 
The IR density was also compatible with that of \citet{aha05}.
Although the electron index of $\gamma_1 =1.2$ is much harder than 
the value of 2 predicted by general acceleration theories 
\citep[e.g.,][]{bla87},
the corresponding radio spectral index
 of $F_\nu \propto \nu ^{\alpha _r}$, $\alpha_r = -0.1$, was
within typical values for PWNe, $-0.3\le\alpha_r\le 0$ \citep{wei78}.
The ratio of observed X-ray to gamma-ray flux showed
the synchrotron cooling was dominant compared to IC cooling. 
Thus, the spectral break was due to the synchrotron cooling,
$E_{\rm br, cool} = 6\pi m_e ^2c^3/(B^2\sigma _T\tau)$,
where $B$ is a magnetic field of the PWN and
$\sigma _T$ is the Thomson cross section.
If we accept a magnetic field of 17\,$\mu$G,
the break in the photon spectrum was expected to occur
at $\sim$0.2\,keV and $\sim$ 1 eV for 
the age of $\tau=1700$\,yr and 20\,kyr, respectively.
However the electron break energy $E_{\rm br}$ of 77 GeV,
derived from the model fit, predicts a break in the photon spectrum
at lower energy of $\sim 2\times 10^{-3}$\,eV,
which suggests larger $B$ or $\tau$.
%
%
%
%
%
By integrating eq.~(\ref{brokenpl}) above 1\,GeV,
the total energy $W_e$ amounts to $3.0\times 10^{48}$\,ergs.
In the case of $\tau = 1700$\,yr and $\tau _0 =30$\,yr,
$\sim$4\% of $E_{\rm tot}$ is required to be converted to $W_e$ ---
an age of 20\,kyr may more easily supply sufficient energy.

\begin{figure}[tp]
\epsscale{}
\plotone{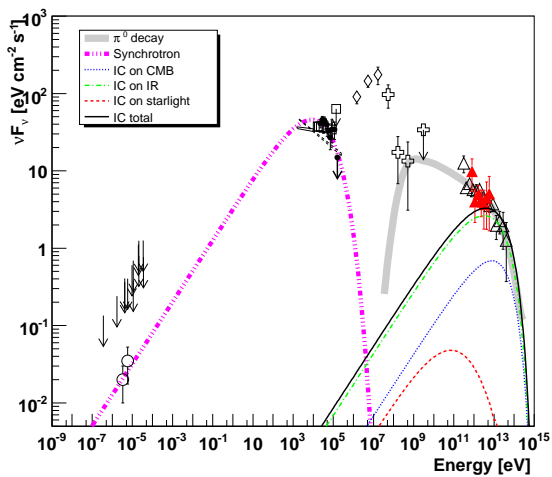}\
\caption{SED and the model calculation of 
  the leptonic scenario with a single power-law electron distribution.
  The representation of each component is the same as in Fig.~\ref{sync1}.}
\label{sync2}
\end{figure}

\noindent
{\bf Case II}:
The expected break in the photon spectrum
at $\sim 0.2$ keV, as mentioned above, 
indicates that the cooling process might not be effective.
Then we could alternatively apply a single power-law electron spectrum using eq.\,(\ref{brokenpl}) with $\gamma _1 = \gamma _2$.
The reproduced SED is shown in Fig.~\ref{sync2}
with parameters in Table~\ref{summary}.
A radio spectral index $\alpha _r=-0.6$,
derived from the electron index of $\gamma_1 =\gamma_2=2.2$,
is close to a mean value for SNRs
of $\alpha _r\sim -0.5$ \citep[e.g.,][]{gre91}
rather than for PWNe of $-0.3\le\alpha_r\le 0$.
%
Here, the IR energy density is more than 3 times higher than that of GALPROP.
There are several possible factors for variations in the local ISRF density
of this magnitude, including uncertainties 
in the model itself but also on the conditions in the environment
of MSH\,15$-$52.
{\it IRAS} 15099$-$5856 is a candidate local
source for an increased IR field, if it is associated with PSR B1509$-$58.
IR from the surrounding dust grains were also suggested by \citet{du95}.
In Fig.~\ref{sync2} the sub-TeV gamma-ray flux was not reproduced.
If we attempt to reproduce it,
the IC peak energy, which was determined by the maximum electron
energy,
must lie below $\sim$0.1 TeV. 
The effect of a stronger magnetic field would be to reduce the IC peaks 
to lower energies and also suppress the peak height.
Since these IC spectra were dominated by the IR component,
if the IR field density is increased, the IC peak is raised vertically in the
SED.
Therefore the combination of a strong magnetic field and IR radiation field,
such as $B\sim$1\,mG and $U_{\rm IR} \sim$7000\,eV/cc,
could reproduce the TeV gamma-ray spectra.
However such a strong magnetic field would 
result in sufficient synchrotron cooling
and cause a photon spectral break at $\sim$0.06\,eV,
in conflict with the assumption of this single power-law model.
Although the requirements from energetics are
looser than that of the broken power-law model,
the SED is not well reproduced.

Finally, in both cases
we can calculate the synchrotron and IC luminosities
from the obtained model curve:
16(case~I)--12(case~II)\% and 0.6--0.4\% of the current spindown energy $\dot{E}(\tau)$
is radiated via synchrotron emission and IC, respectively.
It is also possible to estimate the equipartition
magnetic field strength, $B_{eq}$,
from the fit results as
$B_{\rm eq} = \sqrt{8\pi W_e/V}$, where $V$ is the volume of the emission region.
For Case~I, if we adopt $V=1.3\times 10^{58}$\,cm$^3$ in \citet{sew84},
$B_{\rm eq}=76$\,$\mu$G is obtained, which means the PWN is particle-dominated
as suggested in \citet{che04}.
Assuming $V$ as a sphere at 5.2\,kpc 
with 17$'$ radius (corresponding to the MECS signal region), 
$B_{\rm eq} = 6$\,$\mu$G is obtained, 
which is comparable with the previous indications \citep[e.g.][]{gae02}.
In this case the PWN is poynting-flux dominated.
The Crab nebula is nearly in the equipartition \citep[e.g.,][]{ato99},
while several PWNe have a deviation with a few factor from equipartition \citep{che05}
and for other PWNe the situation is not clear at present.
As for the energetics,
the total energy we have derived,
$W_e$ of $3.0-0.5\times 10^{48}$\,ergs and 
$W_p$ of $3.2\times 10^{50-51}$\,ergs,
depend on the distance and the age of the pulsar.
The uncertainty in the distance would modify the total energy
by only a factor of a few, not an order of magnitude. 
Our estimations have been optimistic
and the energetics would require a higher efficiency $W_e/E_{\rm tot}$
when expansion loss of PWNe is considered \citep{pac73,che92,van01,bej03,che05}.
Although similar studies concerning the source of radiating particles
have recently been started for
other PWNe \citep[eg.,][]{fun07, fun07b},
conclusive solutions have yet to be reached.

\section{Conclusion}

CANGAROO-III observed the SNR MSH\,15-52 containing PSR B1509$-$58 for
48.4 hours in 2006 and detected VHE gamma-ray emission at the 7$\sigma$
level.  The obtained differential flux and the intrinsic extent of
TeV gamma-ray emission are consistent with the previous H.E.S.S.\ result.
Studies of the multi-wavelength spectra was performed, based on both 
hadronic and leptonic models.
In the leptonic scenario, an IR photon field and
a cooled broken power-law spectrum of
electrons are necessary to reproduce the TeV gamma-ray emission.  
From the point of view of the energetics,
if we do not take into account the expansion loss,
a typical supernova could provide sufficient energy for electrons
to reproduce the SED, while hardly to protons.
The morphology of the TeV gamma-ray emission, however,
does not support the supernova explosion as the global energy source. 
Electrons can also be accelerated enough
to reproduce the SED when
$\ge 4$\% ($\tau _0=30$\,yr is assumed) of the rotational energy is
supplied to the kinetic energy.  If the pulsar was older than
its characteristic age of 1700\,yr, e.g., 20\,kyr, the required acceleration
efficiency would be reduced.

Filling in the gaps in the SED is very important for the discussion above.
IR/optical measurements of the PWN emission are crucial 
to determine the synchrotron spectrum and the electron scenario.
Additionally, the determination of the spectral break 
in the synchrotron or IC component would 
help to resolve the long-standing question of the age of this complex system.
We await the results from all sky survey of 
the recently launched IR satellite {\it Akari} \citep{mur07}.
GLAST and the next generation of the ground-based IACTs
such as CTA \citep{cta} and AGIS \citep{agis}
are expected to determine the IC spectra.
In addition, X-ray observations of the whole PWN, 
excluding PSR B1509$-$58,
are required in order to accurately estimate
the flux of synchrotron nebula emission.

\acknowledgements

The authors would like to thank Dr. T.~Mineo for
kindly providing us the reanalysis data of PDS.
We also thank the anonymous referee for helpful comments
to improve the manuscript with a careful reading.
This work was supported by a Grant-in-Aid for Scientific Research by
the Japan Ministry of Education, Culture, Sports, Science and
Technology (MEXT), the Australian Research Council,
and the Inter-University Research Program of the
Institute for Cosmic Ray Research.
The work is also supported by
Grant-in-Aid for 21st century center of excellence programs
``Center for Diversity and Universality in Physics'' 
and ``Quantum Extreme Systems and their Symmetries''
of MEXT.
We thank the Defense Support Center
Woomera and BAE systems and acknowledge all the developers and
collaborators on the GALPROP project.  
T.\ Nakamori and Y.\ Higashi were supported by
Japan Society for the Promotion of Science Research Fellowships 
for Young Scientists.

%
%
%
%
%

%
%
%
%
%
%
%
\begin{deluxetable}{ccccc} 
\tablewidth{0pt}
\tablecaption{Summary of data used in the SED analysis.
\label{seddata}}
\tablehead{
  \colhead{Detector}&
  \colhead{Marker}&
  \colhead{Pulse component}&
  \colhead{Spatial component\protect{\tablenotemark{a}}} &
  \colhead{Reference}
}
\startdata
ATCA                & open circle    & unpulsed    & N& (1)(2) \\
{\it Chandra}/ACIS  & closed region(thick lines)  & pulsed+unpulsed    & NJ& (2)(3) \\
{\it SAX}/MECS      & closed region(thin lines)  & unpulsed & PNJC& (4) \\
{\it SAX}/PDS       & open square    & unpulsed & PNJRCO& (5) \\
{\it INTEGRAL}/IBIS & filled circle  & unpulsed & PNJRC& (6) \\
{\it RXTE}/PCA+HEXTE& closed region(dotted lines)  & unpulsed & PNJRCO& (7) \\
COMPTEL             & open diamond   & pulsed+unpulsed   & PNJRCO& (8) \\
EGRET               & open cross     & pulsed+unpulsed    & PNJRCO& (8) \\
H.E.S.S.            & open triangle  & pulsed+unpulsed    & PNJRC& (9) \\
CANGAROO-III        & filled triangle& pulsed+unpulsed    & PNJRC& this work
\enddata
\tablenotetext{a}{N: Diffuse PWN defined in ref (2),
  J: Jet and outer arc defined in ref (2) and (3), 
  P: Pulsar,  C: Central Diffuse Nebula defined in \citet{tru96},
  R: RCW89,  O: Outside MSH\,15$-$52}
\tablerefs{
  (1) Gaensler et al.\ 1999;
  (2) Gaensler et al.\ 2002;
  (3) Delaney et al.\ 2006;
  (4) Mineo et al.\ 2001;
  (5) Mineo 2007, private communication
  (6) Forot et al.\ 2006;
  (7) Marsden et al.\ 1997;
  (8) Kuiper et al.\ 1999;
  (9) Aharonian et al.\ 2005;
}
\end{deluxetable}

\begin{deluxetable}{ccccccccccc}
\tablewidth{0pt}
\tablecaption{Summary of parameters used in leptonic model.
\label{summary}}
\tablehead{
\colhead{Electron} &
\colhead{$B$} &
\colhead{$\gamma _1$} &
\colhead{$\gamma _2$} &
\colhead{$E_{\rm br}$}&
\colhead{$E_{\rm max}$} &
\colhead{$U_{\rm IR}$} &
\colhead{$W_e$} &
\colhead{$L_{\rm sync}$\protect{\tablenotemark{a}}}&
\colhead{$L_{\rm IC}$\protect{\tablenotemark{b}}} \\
\colhead{spectrum} &
\colhead{[$\mu$G]} &
\colhead{} &
\colhead{} &
\colhead{[GeV]}&
\colhead{[TeV]} &
\colhead{[eV/cc]} &
\colhead{[ergs]} &
\colhead{[ergs/s]} &
\colhead{[ergs/s]} 
}
\startdata
broken P.L.  & 17 & 1.2 & 2.7 & 77& $2.5\times 10^2$ & 2.3 & $3.0\times 10^{48}$ & $2.9\times 10^{36}$ & $1.2\times 10^{35}$ \\
single P.L.  & 20 & 2.2 & 2.2 &    - & $1.3\times 10^2$ & 4.5 & $5.4\times 10^{47}$ & $2.1\times 10^{36}$&$7.7\times 10^{34}$ 
\enddata
\tablenotetext{a}{~Luminosity of synchrotron emission calculated from the model fit curve.}
\tablenotetext{b}{~Luminosity of inverse Compton emission calculated from the model fit curve.}
\end{deluxetable}

\end{document}